\pdfoutput=1
\documentclass[12pt,preprint]{aastex}
\usepackage{graphicx}
\usepackage{natbib}
\usepackage{aas_macros}
\usepackage[section] {placeins}
\usepackage{subfigure}
\usepackage{float}

\input mydefs.sty

\begin{document}

\title{Environmentally Driven Global Evolution of Galaxies}
 
\author{
Renyue Cen$^{1}$
} 
 
\footnotetext[1]{Princeton University Observatory, Princeton, NJ 08544;
 cen@astro.princeton.edu}

\begin{abstract} 

Utilizing {\it high-resolution large-scale} galaxy 
formation simulations of the standard cold dark matter model,
we examine global trends in the evolution of galaxies due to gravitational shock heating
by collapse of large halos and large-scale structure.
We find two major global trends.
(1) The mean specific star formation rate (sSFR) at a given galaxy mass
is a monotonically increasing function with increasing redshift.
(2) The mean sSFR at a given redshift is a monotonically increasing function of decreasing galaxy mass 
that steepens with decreasing redshift.
The general dimming trend with time merely reflects the general decline of 
gas inflow rate with increasing time.
The {\it differential} evolution of galaxies of different masses with redshift 
is a result of gravitational shock heating of gas
due to formation of large halos (groups and clusters) and large-scale structure that 
move a progressively larger fraction of galaxies and their satellites
into environments where gas has too high an entropy to cool to continue feeding resident galaxies.
Overdense regions where larger halos are preferentially located
begin to be heated earlier and have higher temperatures than lower density regions at any given time,
causing sSFR of larger galaxies to fall  
below the general dimming trend at higher redshift than less massive galaxies
and galaxies with high sSFR to gradually shift 
to lower density environments at lower redshift.
We find that several noted cosmic downsizing phenomena are different manifestations of these general trends.
We also find that the great migration of galaxies from
blue cloud to red sequence as well as color-density relation, among others,
may arise naturally in this picture.

\end{abstract}
 
\keywords{Methods: numerical, 
Galaxies: formation,
Galaxies: evolution,
Galaxies: interactions,
intergalactic medium}
 
\section{Introduction}

The intriguing phenomenon of the so-called cosmic downsizing \citep[e.g.,][]{1996Cowie}
has had practioners of the cold dark matter cosmogony perplexed.
Innovative astrophysical ideas have been proposed to 
introduce scales in the growth of galaxies within the context of hierarchical 
formation of dark matter halos 
in the standard cosmological constant-dominated cold dark matter model (LCDM) \citep[][]{2010Komatsu}.
Successful models have been constructed, for example, semi-analytically
by incorporating possible AGN feedback \citep[e.g.,][]{2006Croton,2006Bower}.

In this work we investigate the nature of cosmic downsizing in the LCDM model 
by performing and analyzing {\it high-resolution large-scale} hydrodynamic 
galaxy formation simulations, 
including feedback from star formation and proper treatment
of gravitational heating due to collapse of large-scale structure.
Our simulations reproduce well observations that 
galaxies of higher star formation (SF) rates (SFR) contribute progressively more to the
overall SFR density towards higher redshift
\citep[e.g.,][]{1996Cowie}.
We find that this cosmic downsizing phenomenon is part of 
a fundamental and universal trend that the sSFR, {\it on average}, 
is a monotonic function of galaxy halo (or stellar) mass
with lower-mass galaxies having higher sSFR.
As a result, {\it on average}, 
the stellar mass doubling time is a monotonically decreasing function with decreasing stellar mass
at any redshift and for more massive galaxies that upcrosses the Hubble time 
earlier than less massive galaxies.
The sSFR of galaxies of all masses, {\it on average}, display a monotonic and mass-dependent 
rate of increase with redshift. 
In this sense, we see primarily a trend of ``differential galaxy dimming" from high redshift to $z=0$.
Although the sSFR trend continues to the highest redshift we have examined,
the SFR density that is a convolution of these trends
and halo abundance evolution in the cold dark matter model displays a maximum at $z=1.5-2$. 
Related, within the simulation volume and density fluctuations that we probe, 
we also see an ``upsizing" trend at $z\ge 2$ in that the maximum SFR of galaxies
decreases towards still higher redshift,
probably reflecting the tenet of the standard cold dark matter model
of hierarchical buildup of dark matter halos where the abundance of large, star-forming
halos start to drop off exponentially.

We examine the underlying physical cause for these distinct trends.
We find that at high redshift ($z\ge 2$) SF is largely gas demand limited,
where there is sufficient supply of cold gas for galaxies to double its stellar mass
within a Hubble time and SF is mostly regulated by its own efficiency, 
due to feedback effects from star formation.
At $z\le 2$ SF gradually moves to the regime of being supply limited, dependent on environments,
as the supply rate of cold gas decreases, due to a combination of primarily two factors.
First, the overall decrease of density [$\propto (1+z)^3$] 
causes the gas inflow rate to decline with decreasing redshift.
Second, the overall heating of 
cosmic gas due to formation of large halos (such as groups and clusters) and large-scale 
structures causes a progressively larger fraction of halos to 
inhabit in regions where gas has too high an entropy to cool to 
continue feeding the residing galaxies.
The combined effect is {\it differential} in that overdense regions are heated 
earlier and to higher temperatures than
lower density regions at any given time.
Because larger halos tend to reside, in both a relative and absolute sense,  in more overdense regions than smaller halos,
the net differential effects are that larger galaxies
fall below the general dimming trend at higher redshift 
than less massive galaxies, the sSFR as a function of galaxy mass steepens with time
and galaxies with the high sSFR gradually shift 
to lower density environments.
We do include supernova feedback 
in the simulations and find that galactic winds are strong for
starburst galaxies, strongest
at $z\ge 2$ when SF activities are most vigorous
and are stronger in less massive galaxies than in large galaxies.
But it appears that the stellar feedback processes do not drive any noticeable trend 
of the sort presented here, although  they are important in self-regulating star formation 
at high redshift when gas supply rate is high. 

We also find that the cold gas starvation due to gravitational heating 
provides a natural mechanism to explain the observed migration of galaxies 
to the red sequence from the blue cloud as well as many other phenomena,
such as the observed color-density relation, the trend of galaxies
becoming bluer in lower density environment, and others.

The outline of this paper is as follows.
In \S 2 we detail our simulations, method of making galaxy catalogs and analysis methods. 
Results are presented in \S 3.
In \S 3.1 we compare some basic galaxy observables to observations.
In \S 3.2 we present detailed results and compare to observations.
We then examine and understand physical processes that are primarily responsible for
the results obtained in \S 3.3, followed by predictions of the model in \S 3.4.
Conclusions are given in \S 4.

\section{Simulations}\label{sec: sims}

\subsection{Hydrocode and Simulation Parameters}

We perform cosmological simulations with the adaptive mesh refinement (AMR) 
Eulerian hydro code, Enzo 
\citep[][]{1999aBryan, 1999bBryan, 2004OShea, 2009Joung}.  
First we ran a low resolution simulation with a periodic box of $120~h^{-1}$Mpc on a side.
We identified two regions separately, one centered on
a cluster of mass of $\sim 2\times 10^{14}\msun$
and the other centered on a void region at $z=0$.
We then resimulate each of the two regions separately with high resolution, but embedded
in the outer $120h^{-1}$Mpc box to properly take into account large-scale tidal field
and appropriate boundary conditions at the surface of the refined region.
We name the simulation centered on the cluster ``C" run
and the one centered on the void  ``V" run.
The refined region for ``C" run has a size of $21\times 24\times 20h^{-3}$Mpc$^3$
and that for ``V" run is $31\times 31\times 35h^{-3}$Mpc$^3$.
At their respective volumes, they represent $1.8\sigma$ and $-1.0\sigma$ fluctuations.
The initial condition in the refined region has a mean interparticle-separation of 
$117h^{-1}$kpc comoving, dark matter particle mass of $1.07\times 10^8h^{-1}\msun$.
The refined region is surrounded by two layers (each of $\sim 1h^{-1}$Mpc) 
of buffer zones with 
particle masses successively larger by a factor of $8$ for each layer, 
which then connects with
the outer root grid that has a dark matter particle mass $8^3$ times that in the refined region.
Because we still can not run a very large volume simulation with adequate resolution and physics,
we choose these two runs to represent two opposite environments that possibly bracket the average.
As we have shown in \citet[][]{2010Cen},
these two runs indeed bracket all compared observables of DLAs 
and tests show good numerical convergence.

We choose the mesh refinement criterion such that the resolution is 
always better than $460h^{-1}$pc physical, corresponding to a maximum
mesh refinement level of $11$ at $z=0$.
The simulations include
a metagalactic UV background
\citep[][]{1996Haardt},  
and a model for shielding of UV radiation by neutral hydrogen 
\citep[][]{2005Cen}.
They also include metallicity-dependent radiative cooling 
\citep[][]{1995Cen}.
Star particles are created in cells that satisfy a set of criteria for 
star formation proposed by \citet[][]{1992CenOstriker}.
Each star particle is tagged with its initial mass, creation time, and metallicity; 
star particles typically have masses of $\sim$$10^6\msun$.

Supernova feedback from star formation is modeled following \citet[][]{2005Cen}.
Feedback energy and ejected metal-enriched mass are distributed into 
27 local gas cells centered at the star particle in question, 
weighted by the specific volume of each cell, which is to mimic the physical process of supernova
blastwave propagation that tends to channel energy, momentum and mass into the least dense regions
(with the least resistance and cooling).
We allow the entire feedback processes to be hydrodynamically coupled to surroundings
and subject to relevant physical processes, such as cooling and heating. 
The total amount of explosion kinetic energy from Type II supernovae
for an amount of star formed $M_{*}$
with a Chabrier IMF is $e_{SN} M_* c^2$ (where $c$ is the speed of light)
with  $e_{GSW}=6.6\times 10^{-6}$.
Taking into account the contribution of prompt Type I supernovae,
we use $e_{SN}=1\times 10^{-5}$ in our simulations.
Observations of local starburst galaxies indicate
that nearly all of the star formation produced kinetic energy 
is used to power GSW \citep[e.g.,][]{2001Heckman}.
Supernova feedback is important primarily for regulating star formation
and for transporting energy and metals into the intergalactic medium.
The extremely inhomogeneous metal enrichment process
demands that both metals and energy (and momentum) are correctly modeled so that they
are transported in a physically sound (albeit still approximate 
at the current resolution) way.
The kinematic properties traced by unsaturated metal lines in DLAs are
extremely tough tests of the model, which is shown to agree well with observations \citep[][]{2010Cen}.
As we will show below, the properties of galaxies produced in the simulations 
resemble well observed galaxies, within the limitations of finite resolution.
In order not to mingle too many different effects,
we do not include any feedback effect from AGN, which is often invoked
to suppress star formation by cooling from hot atmosphere in large galaxies.
We will see later that this omission may have caused larger galaxies to be 
somewhat overluminous.

We use the following cosmological parameters that are consistent with 
the WMAP7-normalized \citep[][]{2010Komatsu} LCDM model:
$\Omega_M=0.28$, $\Omega_b=0.046$, $\Omega_{\Lambda}=0.72$, $\sigma_8=0.82$,
$H_0=100 h {\rm km s}^{-1} {\rm Mpc}^{-1} = 70 {\rm km} s^{-1} {\rm Mpc}^{-1}$ and $n=0.96$.


\subsection{Simulated Galaxy Catalogs}

We identify galaxies in our high resolution simulations using the HOP algorithm 
\citep[][]{1999Eisenstein}, operated on the stellar particles, which is tested to be robust
and insensitive to specific choices of concerned parameters within reasonable ranges.
Satellites within a galaxy are clearly identified separately.
The luminosity of each stellar particle at each of the Sloan Digital Sky Survey (SDSS) five bands 
is computed using the GISSEL stellar synthesis code \citep[][]{Bruzual03}, 
by supplying the formation time, metallicity and stellar mass.
Collecting luminosity and other quantities of member stellar particles, gas cells and dark matter 
particles yields
the following physical parameters for each galaxy:
position, velocity, total mass, stellar mass, gas mass, 
mean formation time, 
mean stellar metallicity, mean gas metallicity,
star formation rate,
luminosities in five SDSS bands (and various colors) and others.

For each galaxy we also compute its intermediate-scale environmental overdensity,
defined to be the dark matter density, smoothed by 
a Gaussian function of radius $2h^{-1}$Mpc comoving, divided by 
the global mean dark matter density.
We choose this smoothing scale, because it encloses 
a mass of $1.3\times 10^{13}h^{-1}\msun$, 
whose gas at virial radius shock heated to the virial temperature 
approximately corresponds to the critical entropy $S_{\rm crit}$ that is a weak function of redshift.
The relevance of $S_{\rm crit}$ will be explained in \S 3.2.
In addition, we compute the mean gas entropy of each galaxy at its virial radius,
defined as $<S> = \sum T n^{1/3} dV/\sum n dV$, where the two sums
are over the radial range $(0.9-1.1)r_v$ ($r_v$ is the virial radius).
We also compute various fluxes across the virial radius for each galaxy,
including total gas mass flux, cold mass flux.

\section{Results}

\subsection{Validating Simulated Galaxies}

This is first-in-its-class kind of galaxy formation simulations that includes
sophisticated physical treatment, sufficient resolution, and in a perhaps ground breaking fashion,
a large enough sample covering the entire redshift range 
to statistically address relevant questions.
In \citet[][]{2010Cen} we presented a detailed examination of the DLAs
and found that the simulations, for the first time, are able to match
all observed properties of DLAs, including abundance, size, metallicity and kinematics.
The broad agreement between simulations and observations suggests that our treatment
of feedback processes (including metal enrichment and transport) is realistic;
other simulations that do not include these detailed treatment (such as metal transport) 
do not provide as good agreement with observations as ours especially with respect to kinematics
(that depends quite sensitively on metallicity distribution).
Nevertheless, as with any simulation, there are limitations.
As such, it is prudent to examine the basic properties of galaxies themselves in the simulations 
to gauge how realistically we can reproduce observations.

\begin{figure}[ht]
\centering
\vskip -1cm
\resizebox{5.0in}{!}{\includegraphics[angle=0]{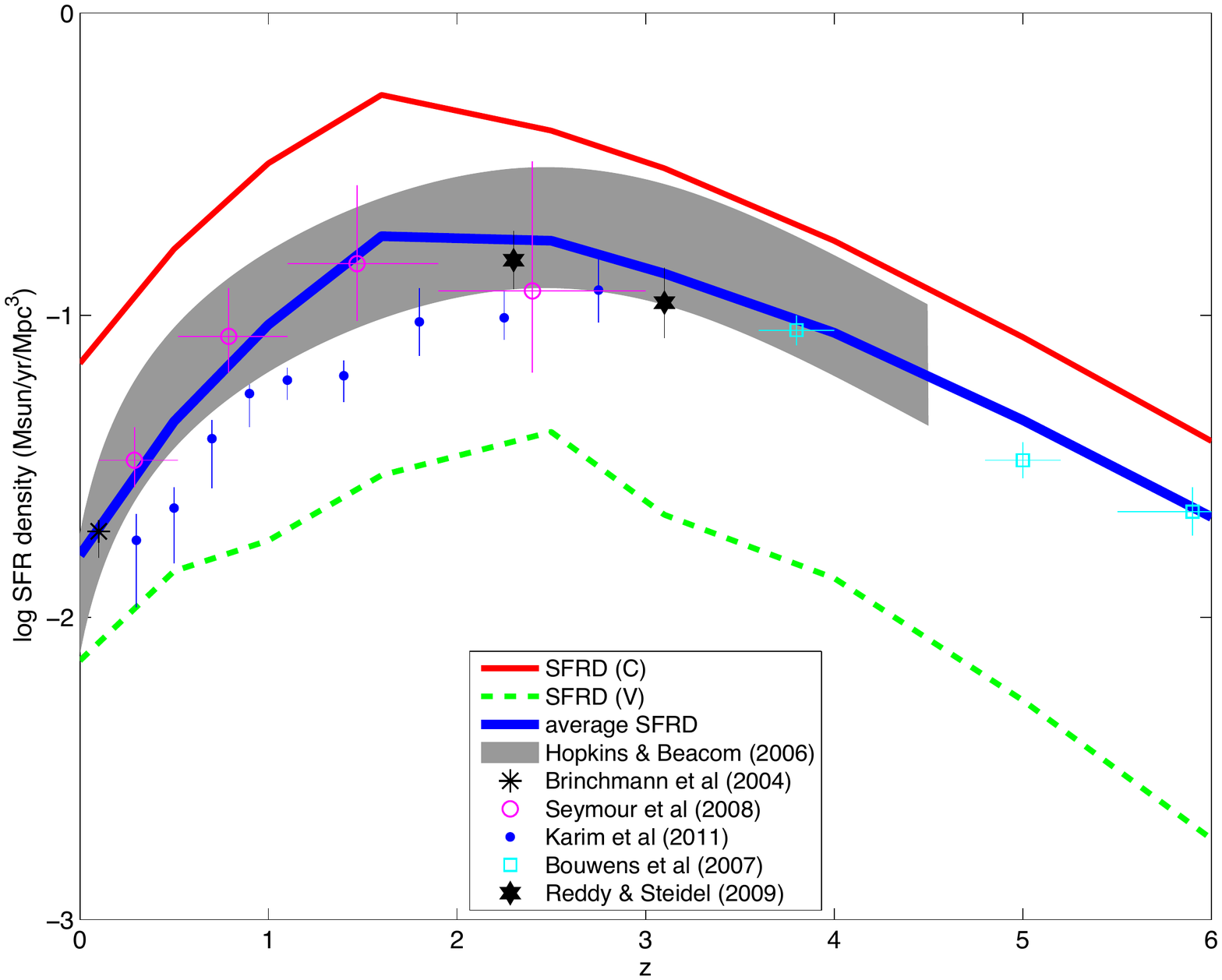}}
\vskip -1cm
\caption{\footnotesize 
shows the evolution SFR density.  
Also shown as the grey shaded region is the observations compiled by 
\citet[][]{2006HopkinsA},
as blue points and mageneta circles two more recent observations using radio techniques 
from \citet[][]{2011Karim} ($2\sigma$ errorbars)
and \citet[][]{2008Seymour} ($1\sigma$ errorbars),
as black asterisk the local SDSS data from 
\citet[][]{2004Brinchmann} ($1\sigma$ errorbars),
as two black hexagons from \citet[][]{2009Reddy} ($1\sigma$ errorbars),
and as open blue squares from \citet[][]{2007Bouwens} ($1\sigma$ errorbars). 
The blue curve is an average of the two runs.
}
\label{fig:SFRz}
\end{figure}

Figure~\ref{fig:SFRz} shows the SFR density history from $z=0$ to $z=6$.
We see that for the entire redshift range  
the SF histories from C and V runs 
bracket the observations, suggesting that the SFR histories in the simulations
are consistent with the observations.
It is probably true that the global average lies between these two runs.
However, the weightings of two runs for averaging 
are likely complicated, because different properties of galaxies of different masses
depend on large-scale environments in a non-trivial fashion.
For brevity, we use the constraints from the observed SFRD history to 
obtain our ``best" weightings for C and V run; 
we find that a weighting for the C run equal to $(1+z)/(7+z)$ (with one minus that for the V run)
to fit the redshift range of interest here,
with the obtained average SFR density shown as the blue curve in 
Figure~\ref{fig:SFRz}.
In some of the subsequent figures, we use the same weightings to average over
some quantities of the two runs, when such an exercise is preferential.

\begin{figure}[ht]
\centering
\vskip -0.5cm
\resizebox{5.0in}{!}{\includegraphics[angle=0]{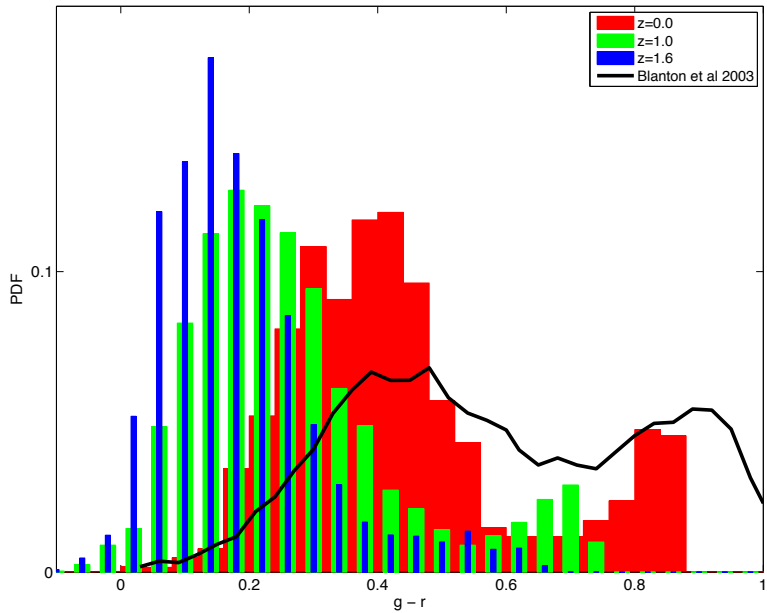}}
\vskip -0.8cm
\caption{\footnotesize 
shows the SDSS restframe $g-r$ color distributions of simulated galaxies (number weighted) 
with stellar mass greater than $10^9\msun$ 
at $z=0,1,1.6$ (red, green and blue, respectively).
Also show as the black curve is the corresponding SDSS observations at $z=0.1$ from 
\citet[][]{2003bBlanton}.
}
\label{fig:bimodal}
\end{figure}

Figure~\ref{fig:bimodal} shows the SDSS restframe $g-r$ color 
distribution of galaxies at $z=0,1.0,1.6$.
The averaged color distribution at each redshift is obtained by the same weighting scheme 
normalized to the SFR density evolution in Figure~\ref{fig:SFRz}.
We see that the simulations can reproduce the observerd bimodality well at $z=0$
\citep[][]{2003bBlanton};
varying the weightings of the two runs in averaging within any reasonable range does not alter the 
bimodal nature of the distribution.
There is a hint that our simulated galaxies may be slightly too blue (by $\sim 0.05$ mag),
which may in part due to the omission of type Ia supernova feedback on a longer time
scale ($\sim 1$Gyr) in the present simulations (we include feedback from SNe II and prompt SNe Ia).
Our future simulations including SNe Ia should verify this.
There is evidence that the color bimodality persists at least to $z\sim 1$ 
but becomes largely absent by $z=1.6$, consistent with observations \citep[e.g.,][]{2005Weiner, 2007Franzetti, 2007Cirusuolo}.

\begin{figure}[ht]
\centering
\vskip -1cm
\resizebox{5.0in}{!}{\includegraphics[angle=0]{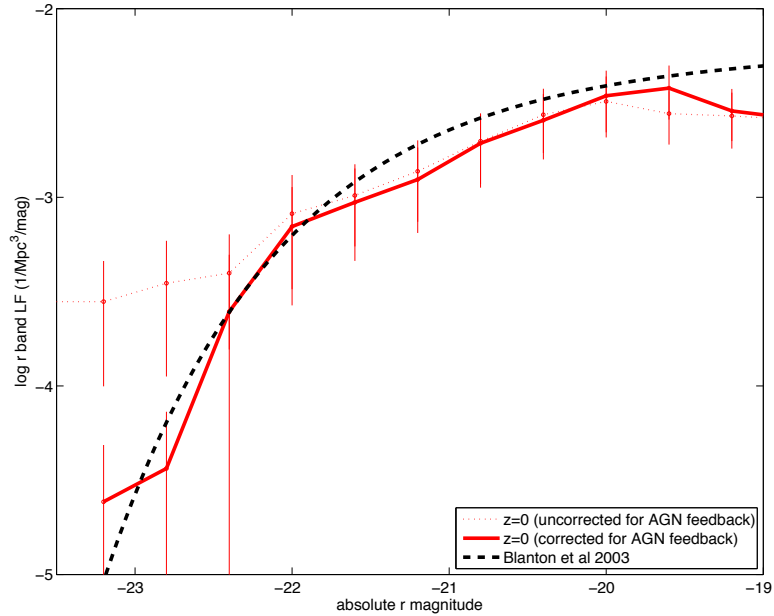}}
\vskip -1cm
\caption{\footnotesize 
shows the SDSS g band galaxy luminosity function at $z=0$. 
The thin dotted curve is directly from averaging over C and V run,
whereas the thick solid curve is obtained after correcting for AGN feedback.
Also shown as the thick dashed curve is the Schechter fit to the SDSS data
\citep[][]{2003Blanton}.
}
\label{fig:LF}
\end{figure}

Figure~\ref{fig:LF} shows the SDSS g band galaxy luminosity function at $z=0$. 
Within the uncertainties the simulations agree reasonably well with observations,
except at the high luminosity end where simulations overproduce luminous galaxies.
This is a well-known problem in simulations that do not include some strong feedback
in large galaxies.
AGN feedback has been invoked
to suppress star formation due to cooling off of hot gas in large galaxies
\citep[e.g.,][]{2006Croton,2006Bower}.
If we apply a similar AGN feedback prescription as in \citet[][]{2006Croton}
by suppressing star formation post-simulation by a factor of 
$f\equiv 1/(1+(M_h/1.0\times 10^{13}\msun)^{2/3})$,
where we use $M_h=M_{star}/0.4$ for satellite galaxies whose halos can no longer
be unambiguously delineated (while stellar identifies remain intact),
we obtain the result shown as the thick solid curve in 
Figure~\ref{fig:LF} that is in good agreement with observations.
There is indication that at $M_g > -19$, we underproduce
small galaxies, which is probably a result of resolution effect.
For the results that we present subsequently, these ``defects" 
do not materially alter any conclusions that we draw, 
because we are mostly interested in evolution of galaxies segregatd in mass
and in environments, which do not depend strongly on precise abundances of galaxies.

\begin{figure}[ht]
\centering
\vskip -1cm
\resizebox{5.0in}{!}{\includegraphics[angle=0]{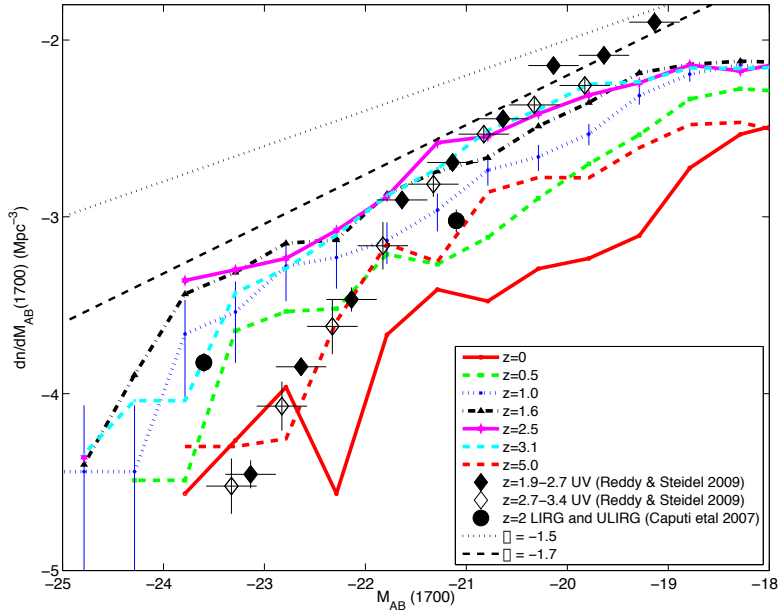}}
\vskip -1cm
\caption{\footnotesize 
shows the rest-frame UV (at $1700\AA$)
luminosity functions at $z=0, 0.5, 1.0, 1.6, 2.5, 3.1, 5$
with $1\sigma$ Poisson errorbars indicated on the $z=1$ curve.
The UV observational data are from \citet[][]{2009Reddy}:
solid diamonds at $z=1.9-2.7$ and open diamonds at $z=2.7-3.4$.
Also shown as two solid dots are observed LIRG and ULIRG data
from \citet[][]{2007Caputi}.
The ULRIG and LIRG data points are shown, if they were not reprocessed through dust,
to account for the fact that we do not process stellar light through dust grains.
The dotted and dashed straight lines indicate the faint end slope of
the luminosity function at $\alpha=-1.5$ and $-1.7$, respectively.
}
\label{fig:SFRF}
\end{figure}

Figure~\ref{fig:SFRF} shows 
the rest-frame UV (at $1700\AA$) luminosity functions at several redshifts,
along with UV and IR (ULIRG and LIRG) observational data,
to check if the reasonable agreement between simulations and observations found at lower redshift 
(Figure~\ref{fig:bimodal} and Figure~\ref{fig:LF}) extend to higher redshifts. 
We convert SFR of each simulated galaxy to $M_{\rm AB}(1700\AA)$ using the standard 
conversion formula, ${\rm SFR} = 6.1\times 10^{-[8+0.4 M_{\rm AB}(1700\AA)]}\msun$/yr
\citep[][]{1998Kennicutt} in combination with the AB magnitude system \citep[][]{1974Oke}.
We see that the simulations agree well with the UV observations 
for $M_{\rm AB}(1700\AA)>-22$, within the uncertainties. 
A significant portion of the disagreement between simulations and UV data at 
$M_{\rm AB}(1700\AA)<-22$ is removed when the abundance of ULIRGs is taken into account,
and the simulations become approximately in agreement with observations within
the errors at $M_{\rm AB}(1700\AA)<-22$.
The faint end slope of the UV luminosity functions appear to be 
steeper than $\alpha=-1.5$ and about $\alpha=-1.8$ to $-1.7$,
consistent with observations \citep[e.g.,][]{2004Yan,2007Bouwens,2009Reddy}. 

In summary, our simulations produce properties of galaxies are in good agreement with
a variety of observations 
that allow us now to examine their global evolutionary trends.

\subsection{Global Trends of Galaxy Formation and Evolution}

\begin{figure}[ht] 
\centering
\vskip -1cm
\resizebox{5.0in}{!}{\includegraphics[angle=0]{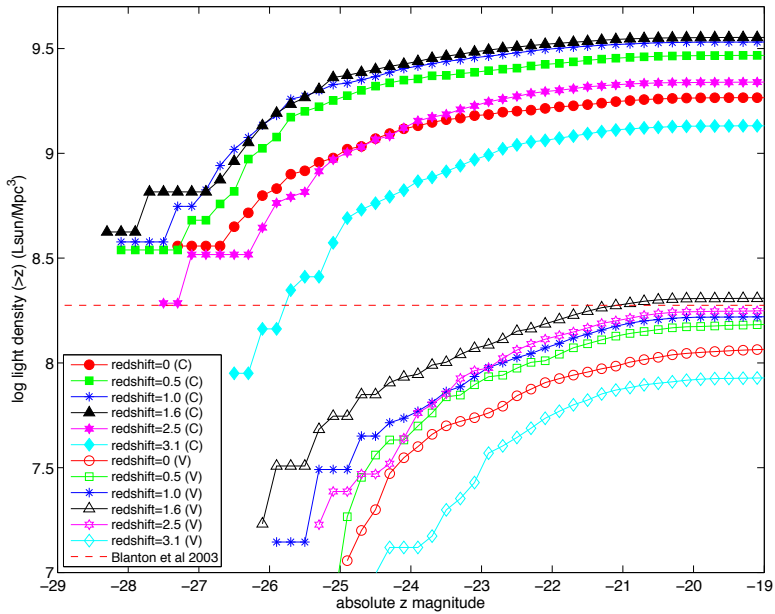}}
\vskip -1cm
\caption{\footnotesize 
shows the cumulative light density distribution in rest-frame SDSS z band as a function of 
absolute z magnitude at redshifts $z=(0,0.5,1.0,1.6,2.5,3.1)$ for both C and V runs.
Also shown as the horizontal dashed line is the value from SDSS data at $z\sim 0.1$
\citep[][]{2003Blanton}.
Similar redshift trends are seen in other SDSS broad bands.
}
\label{fig:Lden}
\end{figure}

\begin{figure}[ht]
\centering
\vskip -1cm
\resizebox{5in}{!}{\includegraphics[angle=0]{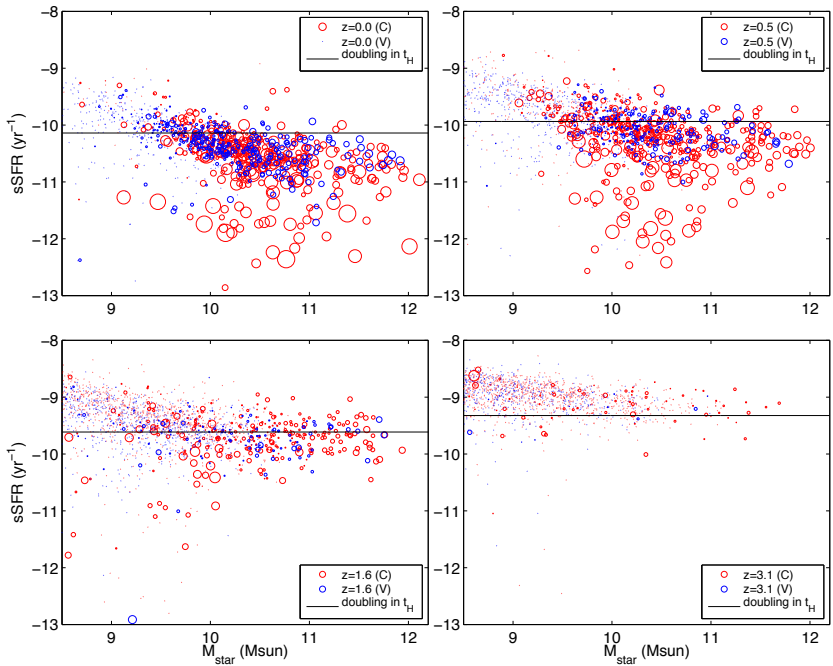}}
\vskip -1cm
\caption{\footnotesize 
shows a scatter plot of sSFR versus galaxy stellar mass
at $z=0$ (top left),
$z=0.5$ (top right),
$z=1.6$ (bottom left) and
$z=3.1$ (bottom right) for both C (red) and V (blue) run. 
Each circle is a galaxy from C (red) and V (blue) run 
with its size proportional to the logarithm of the gas entropy at its virial radius.
The horizontal line in each panel indicates the sSFR value at which a galaxy
would double its stellar mass in a Hubble time.
}
\label{fig:sSFRMstar}
\end{figure}

Figure~\ref{fig:Lden} shows 
the cumulative light density distribution in rest-frame SDSS z band as a function of 
absolute z magnitude from redshift $z=0$ to $z=3.1$.
The fact that the redshift $z=0$ values of the two runs bracket the SDSS data at redshift
$z\sim 0.1$ is self-consistent.
We did not average the two runs in this case, because there is a substantial mismatch 
between the two at $z<-25$, because the abundance of these most luminous galaxies,
at the exponential tail, 
depends more strongly on large-scale environmental density.
We see that from $z=0$ (red circles) to $z=1.6$ (black triangles)
there is a trend that light density increases with increasing redshift, in accord with
the same trend for SFR density seen in Figure~\ref{fig:SFRz}.
It is also seen that the percentage contribution to the light density of galaxies 
at the most luminous end as well as the luminosity of 
the most luminous galaxies increases with increasing redshift from $z=0$ to $z=1.6$.
This particular manifestation is in excellent agreement with the apparent downsizing 
phenomenon first pointed out by 
\citet[][see Figures 6, 20, 24 therein]{1996Cowie}.
As we will show later, the underlying reason for 
this apparent downsizing phenomenon 
is simply that the luminosity function in rest-frame z (or in rest-frame K-band, as shown in 
\citet[][]{1996Cowie})
becomes brighter with increasing redshift from $z=0$ to $z\sim 1.6$,
but the brightening is across the entire spectrum of galaxy masses.
However, the brightening for galaxies of different masses, i.e., sSFR, 
displays an important differential, where
sSFR as a function of stellar mass has a negative slope
that steepens with decreasing redshift,
as shown in Figure~\ref{fig:sSFRMstar} next.

\begin{figure}[ht]
\centering
\vskip -1cm
\resizebox{5in}{!}{\includegraphics[angle=0]{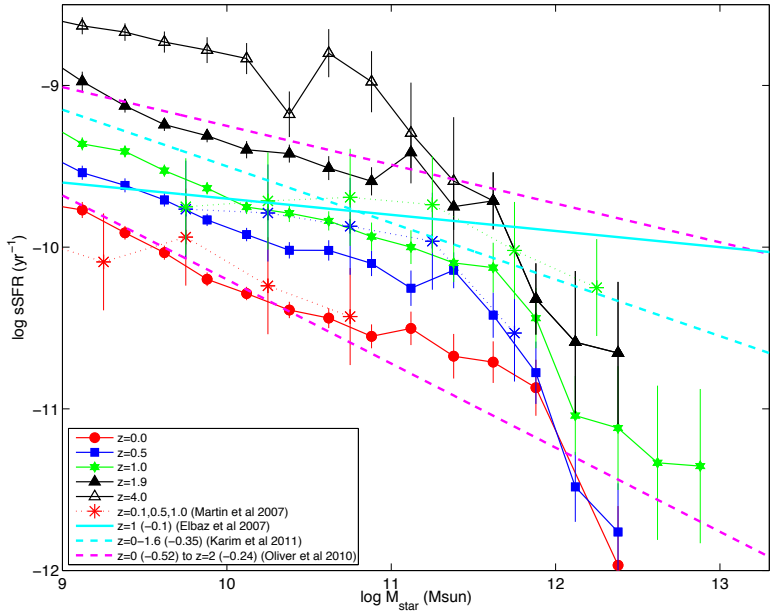}}
\vskip -1cm
\caption{\footnotesize 
shows the average sSFR as a function of stellar mass
at redshifts $z=(0,0.5,1.6,1.9,4.0)$ for both C (solid symbols) and V run (open symbols)
with $1\sigma$ Poisson errorbars.
The IR-to-UV observational data points are from \citet[][]{2007Martin} (red, blue and green asterisks 
for $z=0.1$, $z=0.5$ and $z=1$, respectively) are shown exactly as observed,
from FIR observations of \citet[][]{2007Elbaz} as the cyan line at $z=1$ of $\log SFR - log M_{star}$ slope
of $-0.1$,
from radio observations of \citet[][]{2010Oliver} as two magenta lines are shown 
from $z=0$ of slope $-0.52$ to $z\sim 2$ of slope $-0.24$,
and from radio observations of \citet[][]{2011Karim} as the dashed cyan for the slope range at $z=0-1.6$
of slope $-0.35$.
}
\label{fig:sSFRMstarhis}
\end{figure}

Figure~\ref{fig:sSFRMstar} 
shows the distribution of galaxies
in the sSFR-M$_{\rm star}$ plane at $z=(0,0.5,1.6,3.1)$,
where each galaxy is
also encoded with the average gas entropy at its virial radius - 
a higher entropy corresponds to a larger circle.
The physical importance of gas entropy will become apparent later.
The horizontal line in each panel indicates
the value of sSFR at which the galaxy would double its stellar mass
in one concurrent Hubble time.
We see that at $z=3.1$ (bottom right panel)
most galaxies lie above the horizontal line
and sSFR is nearly independent of stellar mass,
indicating that all galaxies at this redshift are growing at a similar and rapid pace.
As we will show later (see Figure~\ref{fig:flow}),
the cold gas inflow rate significantly exceeds SFR, indicating
that SF is demand based and self-regulated.
Comparison of the four panels clearly shows that 
a progressively larger fraction of galaxies of all masses downcross 
the horizontal line with decreasing redshift,
with larger galaxies starting that migration earlier and generally at a faster pace 
than less massive galaxies.
It is quite visible that the downcrossing of galaxies over the 
horizontal line is accompanied by orders of magnitude increase in gas entropy at
the virial radii of these galaxies, i.e., circles get much larger moving downward.
It is seen that some galaxies of all masses from C run occupy the lower quarter 
of the lower redshift (upper left and upper right) panels
that have the lowest sSFR and largest entropies (large circles);
these are galaxies in high entropy cluster environments.
The negative slope of the sSFR as a function of stellar mass
appears to steepen wth decreasing redshift,
which will be quantified in Figure~\ref{fig:sSFRMstarhis}.
As a result, by $z=0$, only a significant fraction of galaxies
of stellar mass less than $\sim 10^{10}\msun$ can still double their mass
in a Hubble time and they are mostly in the V run (i.e., not in overdense regions), 
while the vast majority of larger galaxies
have lost that ability.
A comparison of red (galaxies from C run) and blue circles (galaxies from V run)
as well as substantial dispersions of sSFR at a fixed stellar mass within each run
indicates that there are substantial variations among galaxies of a same mass
starting at $z=1.6$ that must depend on variables other than just the contemporary galaxy mass.
As will be shown and discussed extensively subsequently,
environmental dependence plays the most fundamental role 
in shaping the formation and evolution of galaxies,
and we find that the gas entropy at the virial radius of each galaxy
is a useful variable for understanding the underlying physical cause.

Figure~\ref{fig:sSFRMstarhis} shows the mean sSFR as a function of stellar mass
at redshifts $z=(0,0.5,1.6,1.9,4.0)$.
We see that simulations show a trend of steepening slope with decreasing redshift,
visually noticed in Figure~\ref{fig:sSFRMstar} above,
which is generally consistent with observations.
The agreement of sSFR between our simulations and IR-UV observations
of \citet[][]{2007Martin} at $z=0-1$ is good within uncertainties.
Currently, the uncertainties in the observed data are still quite substantial, especially at higher redshifts,
as evidenced by 
the differences among the shown observations of \citet[][]{2007Elbaz}, 
\citet[][]{2010Oliver} and \citet[][]{2011Karim} and others (not shown here). 
Nonetheless, there is clear evidence
of a negative slope of sSFR as a function of stellar mass 
that gradually flattens with increasing redshift,
in both our simulations and these observations.

\begin{figure}[ht]
\hskip -0.7cm
\centering
\resizebox{3.71in}{!}{\includegraphics[angle=0]{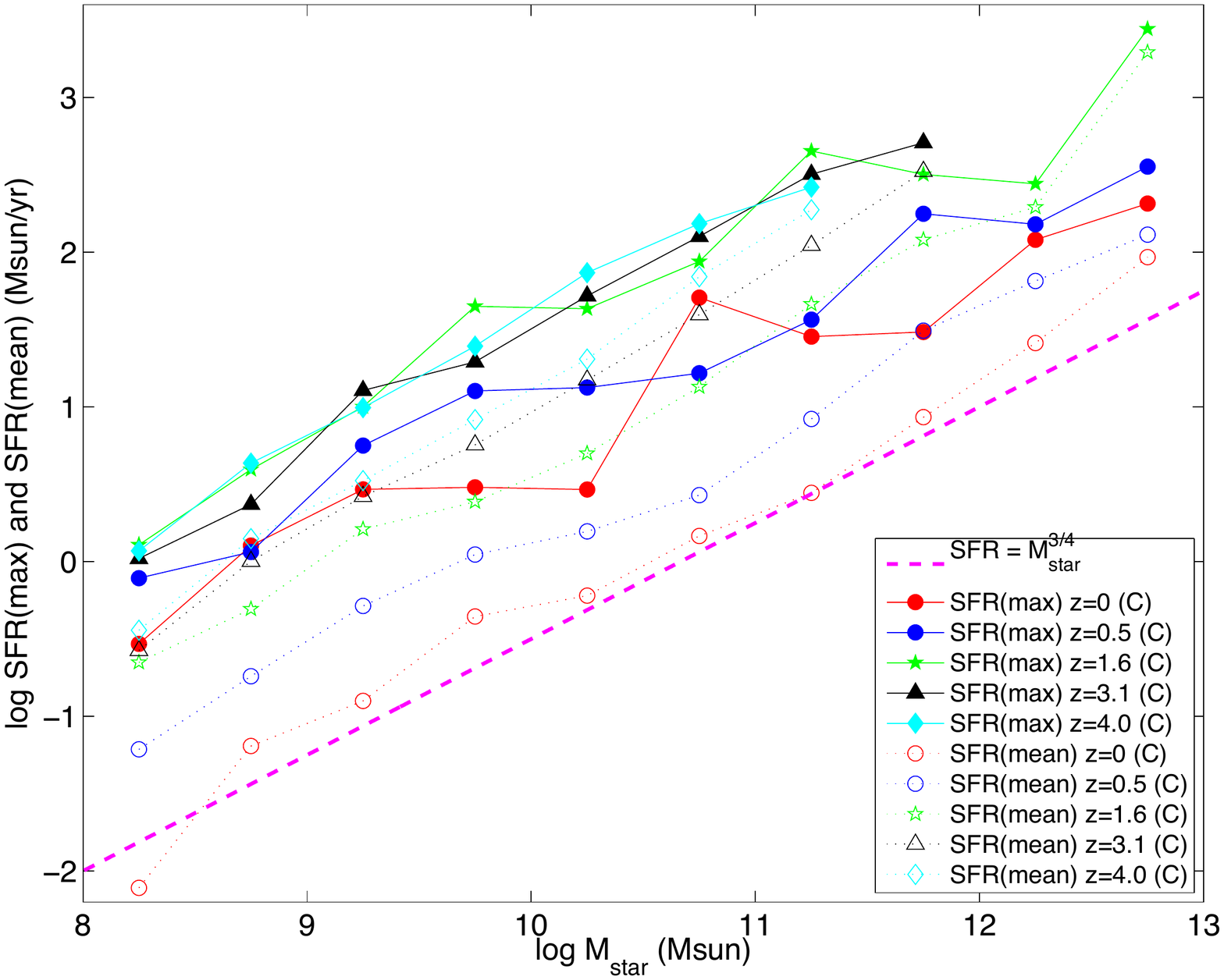}}
\hskip -1.8cm
\resizebox{3.71in}{!}{\includegraphics[angle=0]{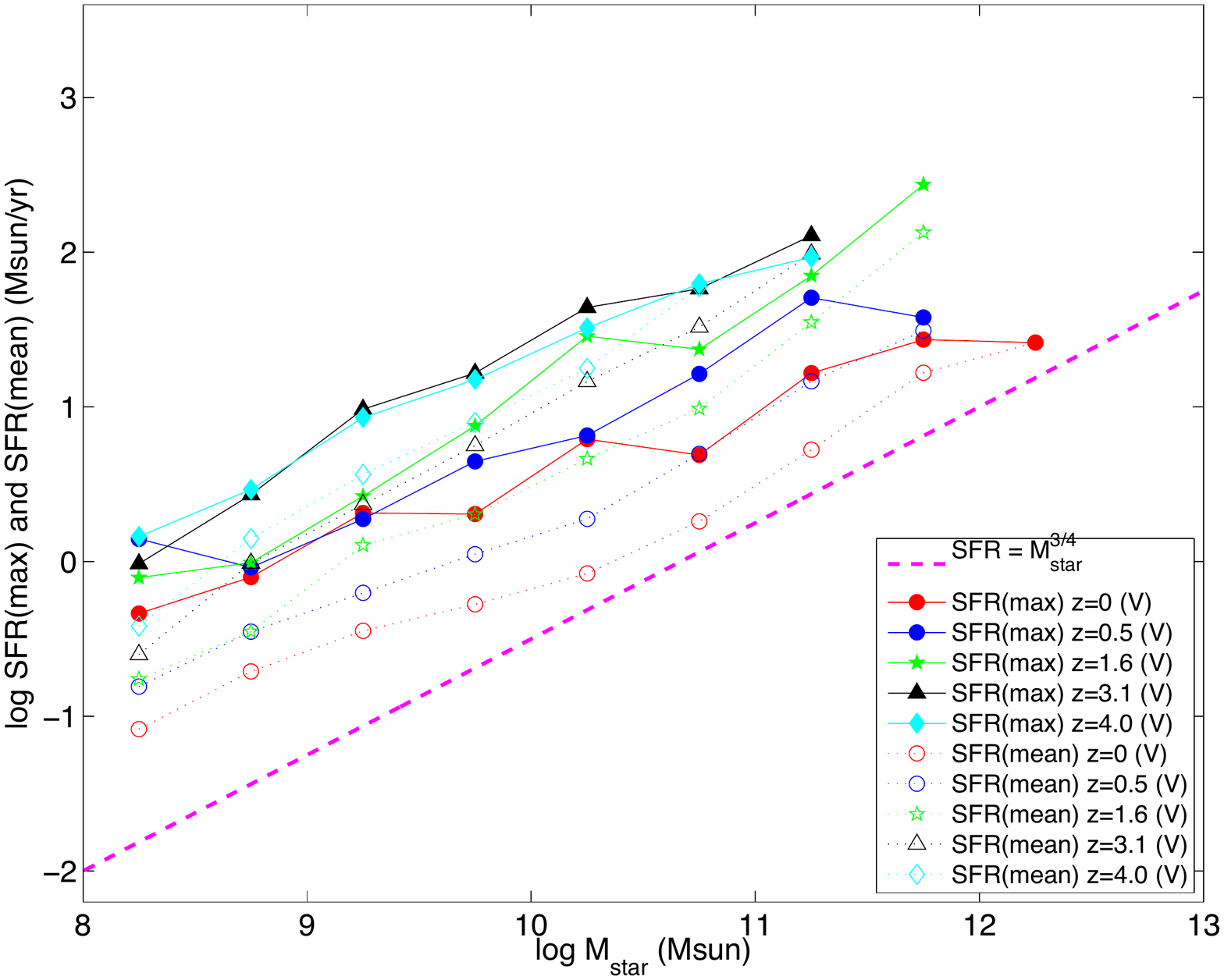}}
\caption{\footnotesize 
shows maximum (solid symbols) and mean SFR (open symbols) 
as a function of stellar mass at redshifts $z=(0,0.5,1.0,1.6,2.5,3.1)$ for both C (left) and V (right) run.
The dashed magenta line has a slope of $3/4$; it is not a fit to the curves but
to guide the eye to see the general trend.
}
\label{fig:SFRmax}
\end{figure}

\begin{figure}[ht]
\centering
\vskip -1.0cm
\resizebox{5.0in}{!}{\includegraphics[angle=0]{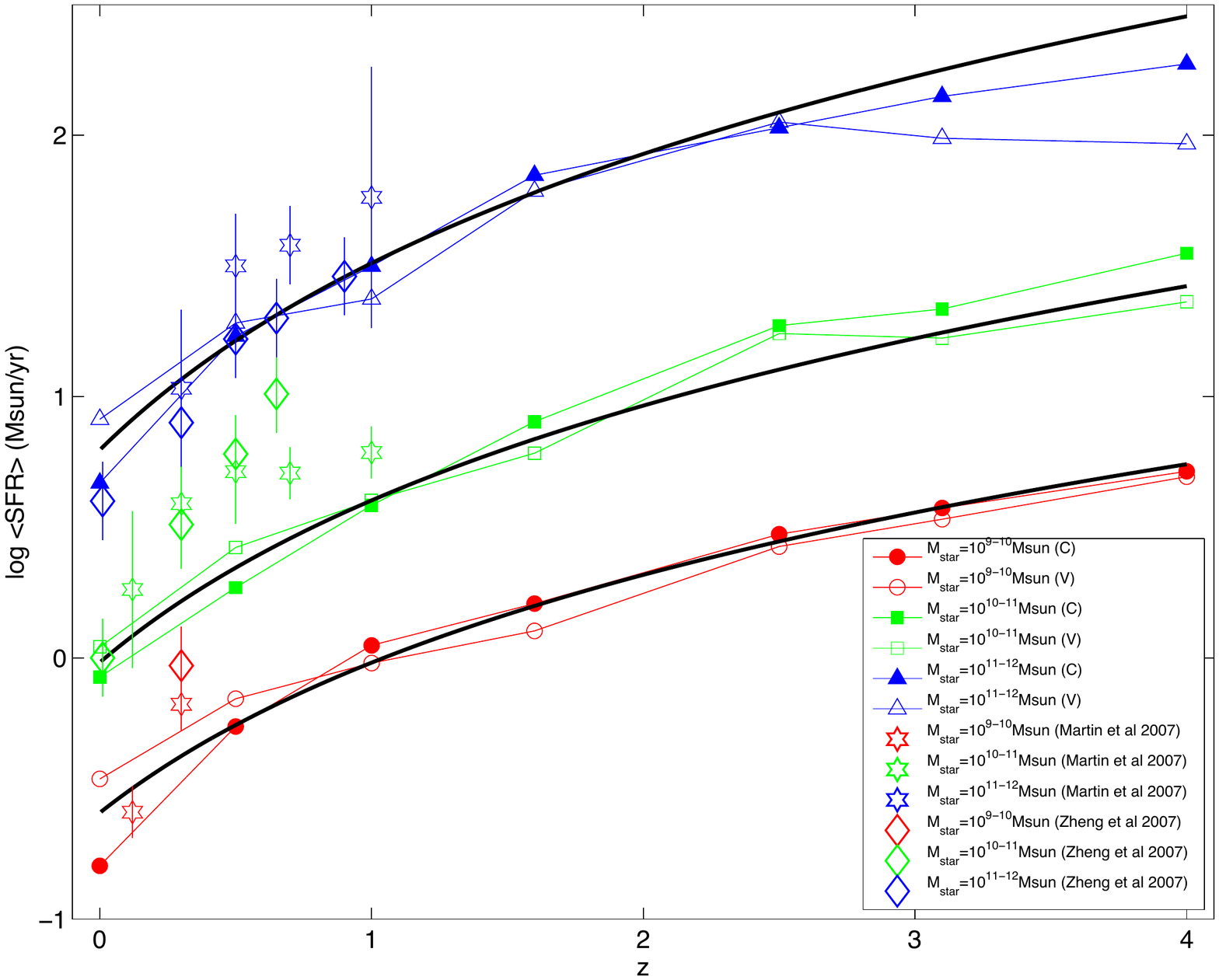}}
\vskip -1.0cm
\caption{\footnotesize 
shows the mean SFR for galaxies of stellar mass in three bins,
$10^{9}-10^{10}$ (red circles), $10^{10}-10^{11}$ (green squares) and $10^{11}-10^{12}\msun$i (blue triangles),
as a function of redshift.
The solid symbols are from C run and open symbols from V run.
The overplotted black curves are 1st order polynomial fits to the three mass bins,
averaged over C and V run curves.
Also shown as hexagons and diamonds are observations from 
\citet[][]{2007Martin} and \citet[][]{2007Zheng}, respectively. 
}
\label{fig:SFRevo}
\end{figure}

In Figure~\ref{fig:SFRmax} we plot the maximum and mean SFR as a function of stellar mass
for seven different redshifts $z=(0, 0.5, 1.0, 1.6, 2.5, 3.1, 4.0)$ for both C (left panel) and V (right panel) runs.
One striking result that is best seen in this plot is that
the maximum SFR of galaxies at a given mass 
increases with increasing redshift 
up to $z_{\rm max}=1.6-3.1$.
Beyond $z_{max}$,
that uptrend for maximum SFR at a fixed mass stops and appears to become static.
Interestingly, the mean SFR at a fixed mass continues to increase
up to the highest redshift shown and the ratio of maximum SFR to mean SFR 
at a fixed mass continues to shrink, reaching a value of $1-3$ in the range $z=2-4$,
suggesting that at high redshift galaxy formation becomes more ``uniform".
The second striking result is that the curves are nearly parallel to one another in the C run,
suggesting that SFR of galaxies of different masses evolve with redshift at similar rates.
This point was noted earlier observationally, first by 
\citet[][]{2007Zheng} (see their Figures 1,2).
As shown in Figure~\ref{fig:sSFRMstarhis},
the rate of change of sSFR for galaxies of different mass galaxies 
is, however, not exactly constant across the mass spectrum.
We see very clearly here by comparing the two panels in 
Figure~\ref{fig:SFRmax}
that this differential at low redshift can be attributed, to a large degree, 
to less massive galaxies in the V run, i.e., in low density environment,
that refuse to join the dimming trend of galaxies in high density environment.
The physical reason for this will be made clear in \S 3.2.

\begin{deluxetable}{cccc}
\tablecolumns{4}
\tablewidth{0pc}
\tablecaption{SFR evolution as a function of stellar mass, 
fitted in the form $\log {\rm SFR}/(\msun {\rm yr}^{-1}) = a (1+z)^b$ -  row 1: stellar mass range; row 2: $a$; row 3: $b$. 
\label{tab:table1}}
\tablehead{
\colhead{Stellar Mass} & \colhead{$10^9-10^{10}\msun$} & \colhead{$10^{10}-10^{11}\msun$} & \colhead{$10^{11}-10^{12}\msun$}}
\startdata
a  & -0.59 & -0.018 & 0.80  \\ 
b  & 1.9 & 2.1 & 2.4
\enddata
\end{deluxetable}

We note that beyond $z_{\rm max}$ the mass at the high end is truncated at progressively smaller values with
increasing redshift.
This sharp cutoff at the high end may be somewhat artificial due to 
the limited simulation box size we have,
but largely reflects the hierarchical nature of growth of dark matter halos
in the standard cold dark matter model.
As we have shown earlier in Figure~\ref{fig:SFRz} and 
Figure~\ref{fig:Lden} the SFR density and light density peak at $z\sim 1.5-2$,
this suggests, in combination with what is seen in Figure~\ref{fig:SFRmax},
that the growth of halos with time dominates over the downsizing trend of SFR down to $z=1.5-2$ from high redshift.
Thereafter, gastrophysical processes that act upon galaxies at $z<1.5-2$ 
cause galaxy formation and evolution to deviate from the track of 
continued hierarchical buildup of dark matter halos, resulting in a trend where
the total luminosity density and SFR density decreases with time and differential evolution 
of galaxies with different masses.

Finally, in Figure~\ref{fig:SFRevo}, we show the redshift evolution of SFR for galaxies in three stellar mass bins:
$10^{9}-10^{10}$ (red circles), $10^{10}-10^{11}$ (green squares) and $10^{11}-10^{12}\msun$ (blue triangles).
The observational data are still relatively uncertain at higher redshift bins for the low-mass galaxies, 
as indicated by the difference between different observational determinations.
The agreement between simulations and observations are reasonable, especially for the highest mass bin.
To best gauge the evolution at low redshift, we decide to fit 
the simulated results using 1st order polynomial fits using only the points at $z<2$,  
although higher (e.g., 2nd) order polynomial fits significantly improve the goodness of the fits at $z\ge 2$.
The best fit parameters are tabulated in Table 1.
It is evident from the fitting parameters that 
higher-mass galaxies suffer a steeper drop in SFR in the range $z=0-2$
than lower-mass galaxies.
This illustrates clearly the differential evolution of sSFR or SFR with redshift for
galaxies of different masses.

\subsection{Physical Origin: Gravitational Heating of External Gas}

We now perform a detailed analysis of the physical conditions 
of galaxies to understand the cause of the trend of cosmic dimming and its differential nature
found in \S 3.2.
A useful starting point may be to quantify the evolution of 
the amount of gas that can cool to feed galaxies.
The amount of gas that can cool depends on density, temperature, metallicity
as well as what happens to the gas subsequently, such as shocks, compression, etc.
It is therefore highly desirable 
to project the multidimensional parameter space to as a low dimension space as possible.
Gas entropy provides an excellent variable to characterize gas cooling properties.
As first insightfully noted by \citet[][]{2004Scannapieco},
the cooling time of any parcel of gas has a minimum 
value that only depends on the entropy of the gas.
Following them we write the gas cooling time in the following form:
\begin{equation}
\label{eq:tcool}
t_{cool} = {(3/2) n k_B T\over n_e^2 \Lambda(T)} = S^{3/2} \left[{3 \over 2}\left({\mu_e\over \mu}\right)^2 {k_B\over T^{1/2}\Lambda(T)}\right],
\end{equation}
\noindent
where $n$ and $n_e$ is total and electron density, respectively;
$k_B$ is the Boltzmann's constant, T temperature and $\Lambda$ cooling function;
$\mu=0.62$ and $\mu_e=1.18$ for ionized gas that we are concerned with;
$S$ is the gas entropy defined as
\begin{equation}
\label{eq:tcool}
S \equiv {T\over n^{2/3}},
\end{equation}
\noindent
in units of K~cm$^2$.
At a fixed $S$ the cooling time is inversely proportional to 
$T^{1/2}\Lambda(T)$. 
The cooling function $\Lambda(T)$ depends on the gas metallicity, which is found in our simulations
to be almost universal at a value of $\sim 0.1\zsun$ for gas at virial radii 
at the redshifts we are interested in here.
Adopting a metallicity of $0.1\zsun$ the term 
$T^{1/2}\Lambda(T)$ has a minimum at $T_{\rm min}\sim 2.3\times 10^5$K
(we note that reasonable variations in metallicity, say, to $0.3\zsun$ from $0.1\zsun$,
does not materially impact our arguments).
Therefore, if $t_{cool}(T_{\rm min}) > t_H$,
the gas can never cool in a Hubble time, because (1) entropy is a non-decreasing 
quantity in the absence of cooling and (2) cooling will be insignificant within $t_H$ 
given the initial requirement.
Subsequent adiabatic compression or expansion does not alter its fate.
Any additional input of entropy, e.g., by shocks,
would increase the entropy and make it more difficult to cool.
Thus, there is a critical value of entropy $S_{\rm crit}$ for any gas above which
gas can no longer cool.
The following fitting formula provides a fit to computed critical entropy $S_{\rm crit}$ 
for gas metallicity of $0.1\zsun$ 
with an accuracy of a few percent over the entire redshift range $z=0-7$: 
\begin{equation}
\label{eq:tcool}
\log [S_{\rm crit}/({\rm K}~{\rm cm}^2)] = 9.183 - 0.167z + 0.0092z^2.
\end{equation}
\noindent

\begin{figure}[ht]
\centering
\vskip -0.5cm
\resizebox{5in}{!}{\includegraphics[angle=0]{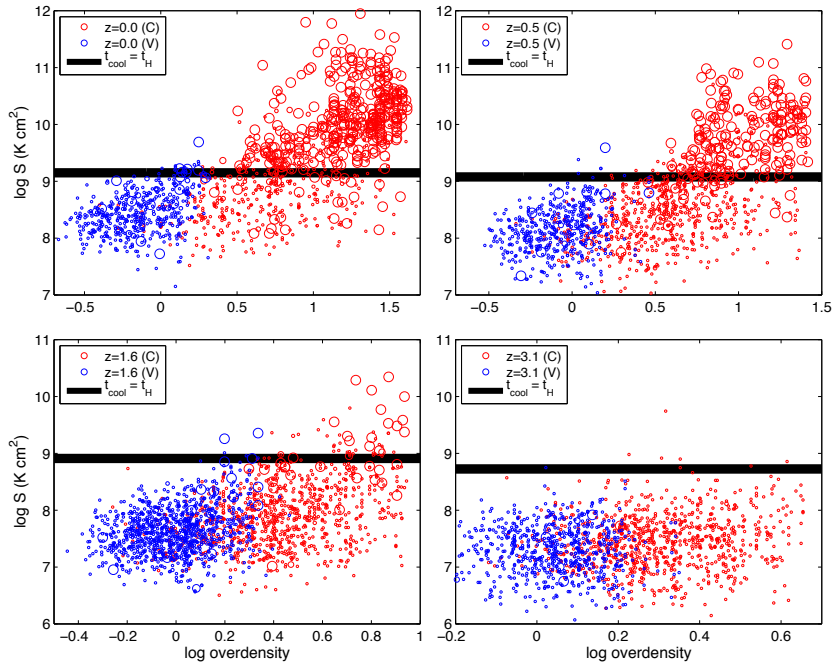}}
\vskip -0.5cm
\caption{\footnotesize 
shows local mean gas entropy at virial radius 
as a function of local overdensity smoothed by a Gaussian window of radius $2h^{-1}$Mpc comoving
at redshifts
$z=0$ (top left),
$z=0.5$ (top right),
$z=1.6$ (bottom left) and
$z=3.1$ (bottom right).
Each circle is a galaxy from C (red) and V (blue) run 
with its size linearly proportional to the inverse of the logarithm of its sSFR;
smaller circles correspond to higher sSFR in this representation.
Also shown as the horizontal bar is the critical entropy $S_{\it crit}$
where cooling time is equal to the Hubble time.
}
\label{fig:entoverd}
\end{figure}

In Figure~\ref{fig:entoverd}
we place each galaxy in the entropy-overdensity parameter plane at four redshifts
($z=0,0.5,1.6,3.1$).
The overdensity is defined to be the dark matter density, smoothed by 
a Gaussian function of radius $2h^{-1}$Mpc comoving, divided by 
the global mean dark matter density.
We see that at $z=3.1$ the entropy of almost all galaxies is located  
below the critical entropy line, indicating that no significant amount
of gas at the virial radius has been heated. 
One should note that, once a gas element has upcrossed 
the critical entropy $S_{\rm crit}$,
it will not fall back below it again.
Therefore, for most galaxies, 
the moment that it upcrosses $S_{\rm crit}$ marks the beginning of the cold gas starvation phase, 
because galaxies tend to move to higher density, higher entropy regions with time.
The size of each circle 
in Figure~\ref{fig:entoverd}
is linearly proportional to the inverse of the logarithm of the sSFR of each galaxy.
We see that galaxies above the $S_{\rm crit}$ line
have dramatically larger circles, i.e., having lower sSFR. 
It is also interesting to see that
galaxies that upcross the $S_{\rm crit}$ line 
do so only in overdense region (smoothed by a Gaussian radius of $2h^{-1}$Mpc).
This is clear and powerful evidence that the differential dimming of galaxies 
is caused by heating of gas in overdense regions;
in other words, galaxy formation and long-term evolution are determined
by external supply of cold gas, 
which in turn depends on overdensity on intermediate scales ($\sim 1$Mpc) 
that dictate the entropy of shock heated gas.

\begin{figure}[ht]
\centering
\vskip -0.5cm
\resizebox{5in}{!}{\includegraphics[angle=0]{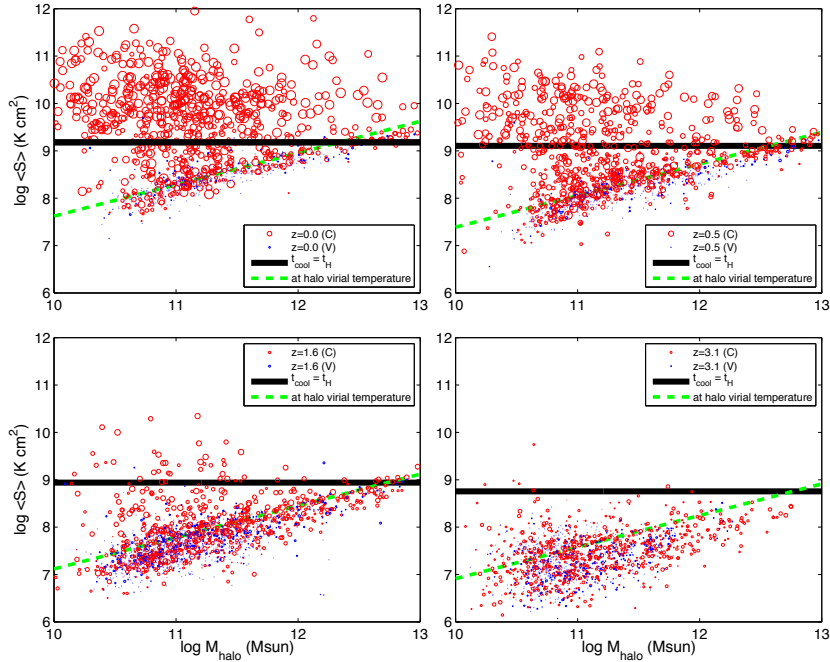}}
\vskip -0.5cm
\caption{\footnotesize 
shows local mean gas entropy at virial radius 
as a function of halo mass at redshifts
$z=0$ (top left),
$z=0.5$ (top right),
$z=1.6$ (bottom left) and
$z=3.1$ (bottom right).
Each circle is a galaxy from C (red) and V (blue) run
with its size proportional to the logarithm of the local overdensity 
smoothed by a Gaussian window of radius $0.5h^{-1}$Mpc comoving.
Also shown as the horizontal bar is the critical entropy $S_{\it crit}$
where cooling time is equal to the Hubble time.
The inclined line indicates the gas entropy at virial radius
if the temperature is exactly equal to the virial temperature of the halo.
}
\label{fig:entMhalo}
\end{figure}

To help further understant this, in Figure~\ref{fig:entMhalo},
we plot the galaxies in the 
entropy-halo mass parameter plane at four redshifts.
Also shown as the dashed green line in each panel is the gas entropy 
at virial radius, if the temperature is heated up to the virial
temperature of the host halo itself. 
One notices that at $z=3.1$ when no galaxies more massive than $\sim 5\times 10^{12}\msun$
has formed, virial heating due to formation of halos is insufficient 
to upcross the entropy barrier.
This is the redshift range where an ample amount of cold gas is available to feed
galaxy formation, resulting in sSFR that is very weakly mass-dependent
and galaxy formation in the ``upsizing" domain,
in concert with the hierarchical buildup of dark matter halos.

At lower redshifts, formation of larger halos
more massive than $\sim 1\times 10^{13}\msun$ (i.e., groups and clusters) 
as well as collapse of larger waves due to formation of large-scale structures (filaments and walls)
 raise a progressively larger fraction of regions to higher entropy than $S_{\rm crit}$.
This causes a dichotomy in the entropy distribution, especially at the low halo mass end ($\le 10^{11}\msun$)
as follows.
There is a branch of low-mass galaxies in low density environments,
as evidenced by their small circle sizes,
which are located along or below the green line in Figure~\ref{fig:entMhalo}
and have entropies comparable to or lower than what is produced due to adiabatic shock heating 
accompanying the formation of the halos themselves.
These small galaxies correspond to galaxies in the upper left corner 
in Figure~\ref{fig:sSFRMstar} that are still able to double their mass in a Hubble time.
Then there is another branch of small galaxies that lie above 
the $S_{\rm crit}$ line and are in overdense regions, 
as evidenced by their large circle sizes.
These small galaxies are red and dead, correspond to dwarf galaxies in heated filaments and group/cluster environments.
Generally, the gas entropy of galaxies above the green dashed line 
is higher than what virial shock heating due to the formation of the halo itself produces;
therefore, all these galaxies above the green line are in essence 
``satellite" galaxies within a large halo (such as a group or cluster) or, if one were to generalize it,
``satellite" galaxies in a gravitational shock heated region due to collapse of large-scale
structure (filaments or pancakes), not necessarily virialized.
The concentration of galaxies with entropy along the green line is due to 
virial shock heating of halo itself, i.e., the primary galaxy.
It is striking that even at $z=0$ there is only a very handful of 
(blue circle) galaxies with mass greater than $10^{12}\msun$
that lie above the $S_{\rm crit}$ from the V run.
Taken together, this is unequivocal evidence that it is the external gas heating
that drives the gas supply hence star formation and galaxy evolution;
the absence of such heating in the V run has allowed galaxies there 
to remain active in star formation at present.

\begin{figure}[ht]
\centering
\vskip -0.5cm
\resizebox{5in}{!}{\includegraphics[angle=0]{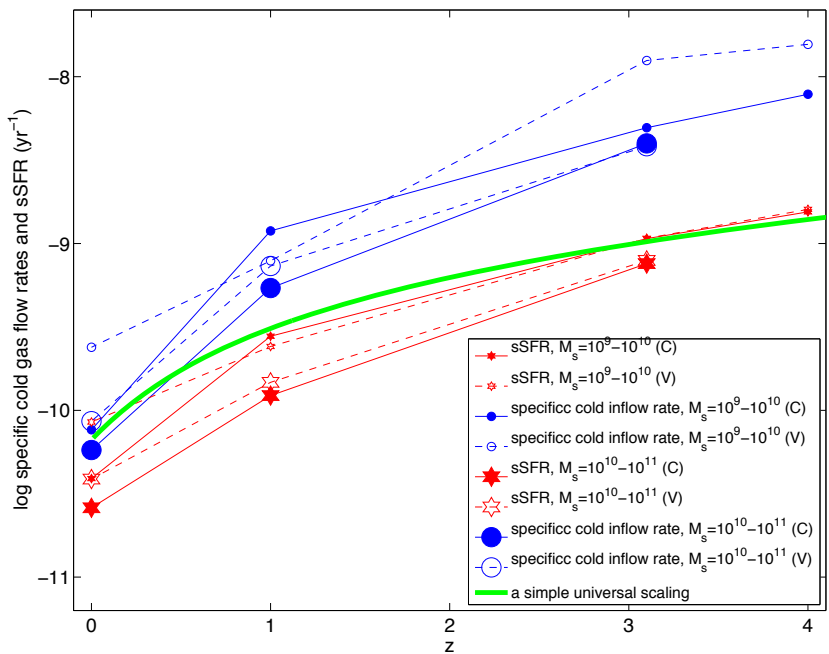}}
\vskip -0.5cm
\caption{\footnotesize 
shows the mean specific cold gas inflow rate (defined to be the cold gas inflow rate 
per unit stellar mass) and mean sSFR for galaxies in two different stellar mass bins for C and V run. 
The cold gas is defined to be that with cooling time less than the dynamic time of the galaxy.
Also shown as solid green curve is the general scaling of gas inflow rate,
which is assumed to be proportional to $4\pi r_v^2(z) v_v(z) \rho^3(z)$, where $r_v$, $v_v$
and $\rho(z)$ are redshift-dependent virial radius, virial velocity and mean gas density. 
}
\label{fig:flow}
\end{figure}

Figure~\ref{fig:flow} shows the mean specific cold gas inflow rate (defined to be the cold gas inflow rate 
per unit stellar mass and cold gas is defined to be gas that has a cooling time less than
the galaxy dynamical time at the virial radius) and mean sSFR for galaxies in two different stellar mass bins
for C and V run. 
Several points are worth noting.
First, we see that the cold gas inflow rates are generally higher than 
star formation rates, suggesting self-regulation of star formation, mostly due to 
feedback from star formation.
Second, the ratio of cold gas inflow rate to SFR decreases with decreasing redshift,
pointing to a gradual transition of SF regimes from gas demand based at high redshift to gas supply
based at low redshift.
Third, the rough similarity between the evolution of the gas inflow rate
based on a simple scaling and the actual computed rates suggests
the bulk of the cosmic dimming trend with decreasing redshift
can be attributed to the decrease of mean density of the universe with increasing time
and the evolution of the Hubble constant (or density parameter).
Finally, the gravitational heating effects add a differentiating process
on top of this general dimming trend, evident here by the different steepening with decreasing redshift
of the specific gas inflow rates and SFR at lower redshifts among
galaxies of different masses and galaxies in different environments (C versus V run).

\begin{figure}[ht]
\centering
\vskip -0.5cm
\resizebox{5in}{!}{\includegraphics[angle=0]{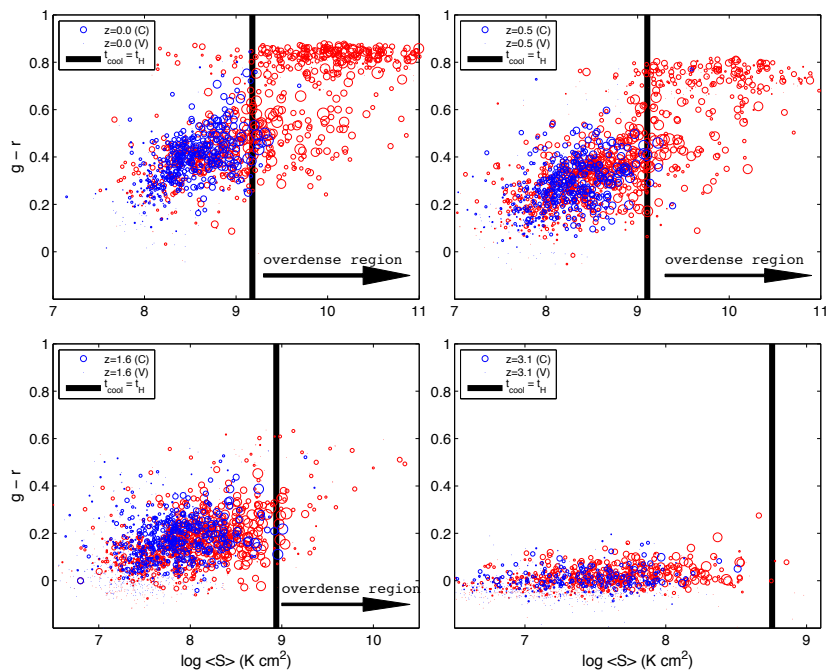}}
\vskip -0.5cm
\caption{\footnotesize 
shows SDSS g-r color of galaxies as a function of local mean gas entropy at the virial radius 
at $z=0$ (top left),
$z=0.5$ (top right),
$z=1.6$ (bottom left) and
$z=3.1$ (bottom right).
The galaxies in C run are shown in red and those in V run in blue.
The size of each circle is proportional to the logarithm of the galaxy stellar mass.
Also shown as the vertical line is the critical entropy $S_{\it crit}$
where cooling time is equal to the Hubble time.
}
\label{fig:GRent}
\end{figure}

Finally, in Figure~\ref{fig:GRent},
we place galaxies in the color-entropy plane. 
Four things are immediately noticeable.
First, the vast majority of galaxies are blue (in color, not the color of the plotted circles) and there is 
no strong evidence of bimodality in color at $z\ge 1.6$.
Second, at $z=0-0.5$,
almost all galaxies in the V run occupy the blue peak at $g-r\sim 0.2-0.6$ with very few in the red peak.
Third, the vast majority of galaxies on the left side of the critical entropy line
are in the blue cloud, as they should.
Fourth, there is a significant number of galaxies on the right side of the critical entropy line
that appear blue and have masses covering a comparable range compared to those
in the red sequence.
Thus, Figure~\ref{fig:GRent} gives a physical underpinning 
for the well-known color-magnitude diagram of galaxies \citep[e.g.,][]{2004Baldry}. 
The existence of the cold-gas-starved yet blue galaxies
indicates that external gas heating is the driving force to cause
these blue galaxies to migrate upward in 
Figure~\ref{fig:GRent} to ultimately join the red sequence.
The fact that many galaxies in the V run, although having higher sSFR than those in the C run 
(see Figures ~\ref{fig:sSFRMstar},  Figures ~\ref{fig:SFRevo}, Figure~\ref{fig:entoverd} and Figure~\ref{fig:flow}),
both remain blue (as they should, given the high sSFR) and have low entropies
suggest that SF is not the primary driver for the color migration.
Internal driver, such as feedback from starbursts or AGN, may play
a role in quenching star formation in a small fraction of galaxies that experience immense starbursts
(e.g., caused by major mergers);
but the situation is unclear at present.

\subsection{Predictions}

Several manifestations of downsizing trends 
should by now be understood, including 
(1) the epoch of major stellar mass buildup in massive galaxies 
is substantially earlier than the epoch of mass buildup in low-mass galaxies, 
(2) the SF and stellar mass buildup are accelerated in overdense regions compared to less overdense regions,
(3) massive galaxies are on average older than less massive galaxies,
(4) galaxies of all masses, on average, get bluer with increasing redshift,
(5) galaxy self metal enrichment shifts from high-mass galaxies at high redshift to lower-mass galaxies at lower redshift,
all in broad agreement with a variety of observations
\citep[e.g.,][]{2004Kodama, 2005PerezGonzalez, 2006Bundy, 2007Noeske, 2007Zheng, 2007Martin, 2007Tresse, 2007Buat, 
2008Lehmer, 2009Mobasher, 2010Hartley, 2010Cirasuolo, 2011Karim, 2011Pilyugin}.
This model provide a coherent and unified physical interpretation.

Many other general trends in galaxy formation and evolution that this model
would predict have already been confirmed by observations,
including 
(1) the galaxy color-environment relation \citep[e.g.,][]{2005Blanton},
(2) galaxy star formation as a function of environment, specifically the dramatic transition
at a few cluster virial radii that mark location of virial shocks \citep[e.g.,][]{2003Gomez},
(3) the trend of galaxies having higher sSFR and becoming bluer towards voids from cluster environments
\citep[e.g.,][]{2004Kauffmann, 2004Rojas, 2005Rojas},
(4) redder galaxies have stronger correlation functions than blue galaxies,
irrespective of their luminosities \citep[e.g.,][]{2005Zehavi}.

Several additional relatively robust trends may be predicted:
(1) the faint end slope of the galaxy luminosity function should approach the Press-Schechter value
of $\sim -2.0$ at high redshift $z\ge 6$, subject to uncertain effects of cosmological reionization.
(2) Cross-correlation between CMB Sunyaev-Zeldovich $y$ maps and density of red galaxies
is expected to be positive and the opposite is true for that between $y$ maps and density of blue galaxies.
(3) Correlations (galaxy-galaxy lensing) 
between background galaxy shapes and foreground red galaxies 
should be systematically stronger than between background galaxy shapes and foreground blue galaxies.

\section{Conclusions}

With high resolution and a physically sound treatment of relevant physical processes,
our state-of-the-art, adaptive mesh-refinement Eulerian cosmological hydrodynamic simulations 
reproduce reasonably well some key observables of galaxies as a whole,
including luminosity function, color distribution and star formation history. 
This allows us to examine, in addition, with confidence, 
some global trends of formation and evolution of galaxies.  
Several findings are interesting and new.

(1) The overall dimming trend of galaxies of all masses is largely attributable to  
the evolution of mean cosmic gas density and density parameter.
(2) Gravitational shock heating due to formation of halos and large-scale structure
adds a differential layer on top of this general global dimming trend.
(3) As a result, 
the mean sSFR is a monotonically increasing function of redshift at a given galaxy mass.
(4) The mean sSFR is a monotonically decreasing function of galaxy mass at a given redshift
and steepens with decreasing redshift,
which overwhelmed the continued hierarchical growth of halos at low redshift range $z=0-2$
and is the underlying physical driver for some apparent ``anti-hierarchical" manifestations of some 
galaxy properties.
(5) The SFR function is a convolution of sSFR and galaxy mass function
 - it increases from $z=0$ to $z\sim 2$ and thereafter decreases towards (i.e., an upsizing trend) higher redshift.
(6) Although the buildup of dark matter mass and stellar mass are not necessarily exactly
parallel to one another, the overall trend for both is still hierarchical.

The underlying physical cause Trend (2) above is as follows.
With time, more regions are heated to higher temperatures
due to formation of large halos (such as groups and clusters) and large-scale 
structures that result in a progressively larger fraction of halos 
inhabiting in regions where gas has too high an entropy to cool to 
continue feeding the residing galaxies.
Thus, overdense regions enter the cold gas starvation phase
earlier than lower density regions.
Because larger halos tend to reside in more overdense regions than smaller halos,
the net differential effects are that larger galaxies
fall below the general dimming trend at higher redshift 
than less massive galaxies, the sSFR as a function of galaxy mass steepens with time
and galaxies with the high sSFR gradually shift 
to lower density environments.
By $z=0$, galaxies with high sSFR (such that they may be categorized as blue) 
have almost entirely left the cluster environments 
and can be found in fields and voids.
Thus, the processes that drive galaxy evolution 
are mostly external at $z\le 2$, due to gravitational heating of either its own halo 
formation, 
or formation of the primary galaxy or group/cluster halo in the case of a satellite galaxy,
or collapse of embedding large-scale structures such as filaments or Zeldovich pancakes,
which at low redshift correspond to the cosmic web of warm-hot intergalactic medium \citep[e.g.,][]{1999Cen}.

We also find that the cold gas starvation due to gravitational heating 
provides a viable physical mechanism to explain the observed migration of galaxies 
to the red sequence from the blue cloud as well as many other phenomena,
such as the observed color-density relation, the trend of galaxies
becoming bluer in lower density environment, and others. 
Several predictions are made in \S 3.4.

As a site note, these findings may imply that the concept of two modes of gas accretion onto galaxies
\citep[e.g.,][]{2005Keres,2006Dekel},
while very useful to crystallize some aspects of galaxy formation, 
may need to be mended to be globally applicable,
because the amount of cold as well as hot gas around a galaxy depends on both 
its mass and its external environment (and perhaps its own history).
For example, a small galaxy in a cluster environment would have a very different 
mix of cold and hot gas components from a galaxy of the same mass in a void environment,
with the latter having a much larger cold gas fraction than the former.
We further note that galaxy formation recipes, such as those used in semi-analytic modeling,
may need to include the important external effects found here to be physically realistic.
In essence, realistic treatments of galaxy formation have to be multivariant,
not just dependent on the contemporary halo mass.

\vskip 1cm

I would like to thank Dr. M.K.R. Joung for help on
generating initial conditions for the simulations and running a portion
of the simulations and Greg Bryan and John Wise for help with Enzo code.
I would like to thank Dr. D. Christopher Martin for kindly
providing plotting data for observations.
Computing resources were in part provided by the NASA High-
End Computing (HEC) Program through the NASA Advanced
Supercomputing (NAS) Division at Ames Research Center.
This work is supported in part by grants NNX08AH31G and NAS8-03060. 
The simulation data are available from the author upon request.


\end{document}